\documentclass[aps,prx,reprint, twocolumn,letterpaper,superscriptaddress,10pt]{revtex4-2}
\usepackage{amssymb,amsthm,amsmath,amsfonts,braket, physics,algorithm2e, float}
\usepackage{graphicx, soul, enumerate}
\usepackage[normalem]{ulem}
\usepackage[justification=Justified]{caption}
\usepackage{subcaption}
\usepackage{comment}
\usepackage{siunitx}
\usepackage[version=4]{mhchem}
\usepackage[a4paper, total={7in, 9in}]{geometry}
\usepackage[colorlinks=true,urlcolor=black,citecolor=black,linkcolor=black]{hyperref} 

\begin{document}
\title{Quantum Key Distribution with Entanglement-Swapped Photons from a Quantum Emitter} 

\date{\today}
\author{Michele~B.~Rota}
\email{michele.rota@uniroma1.it}
    \affiliation{Dipartimento di Fisica, Sapienza Università di Roma, Piazzale Aldo Moro 5, 00185 Roma, Italy}
  
\author{Francesco~Basso~Basset}
    \affiliation{Dipartimento di Fisica, Sapienza Università di Roma, Piazzale Aldo Moro 5, 00185 Roma, Italy}
    \affiliation{Dipartimento di Fisica, Politecnico di Milano, Piazza Leonardo da Vinci 32, 20133 Milano, Italy}
    
\author{Alessandro~Laneve}
    \affiliation{Dipartimento di Fisica, Sapienza Università di Roma, Piazzale Aldo Moro 5, 00185 Roma, Italy}
    
\author{Francesco~Salusti}
    \affiliation{Institute for Photonic Quantum Systems (PhoQS), Center for Optoelectronics and Photonics Paderborn (CeOPP) and Department of Physics, Paderborn University, Warburger Straße 100, 33098, Paderborn, Germany}
    
\author{Nicolas~Claro-Rodriguez}
    \affiliation{Institute for Photonic Quantum Systems (PhoQS), Center for Optoelectronics and Photonics Paderborn (CeOPP) and Department of Physics, Paderborn University, Warburger Straße 100, 33098, Paderborn, Germany}

    \author{Giuseppe~Ronco}
    \affiliation{Dipartimento di Fisica, Sapienza Università di Roma, Piazzale Aldo Moro 5, 00185 Roma, Italy}
    
\author{Mattia~Beccaceci}
    \affiliation{Dipartimento di Fisica, Sapienza Università di Roma, Piazzale Aldo Moro 5, 00185 Roma, Italy}

\author{Tobias~M.~Krieger}
    \affiliation{Institute of Semiconductor and Solid State Physics, Johannes Kepler University, Altenbergerstraße 69, Linz 4040, Austria}

\author{Quirin~Buchinger}
    \affiliation{Technische Physik, University of Würzburg, Am Hubland, D-97074 Würzburg, Germany}

\author{Saimon~F.~Covre~da~Silva}
    \affiliation{Institute of Semiconductor and Solid State Physics, Johannes Kepler University, Altenbergerstraße 69, Linz 4040, Austria}
    \affiliation{Universidade Estadual de Campinas, Instituto de F\'{i}sica Gleb Wataghin, 13083-859 Campinas, Brazil}
 
\author{Sandra~Stroj}
    \affiliation{Research Center for Microtechnology, Vorarlberg University of Applied Sciences, CAMPUS V, Hochschulstraße 1, 6850, Dornbirn, Austria}

\author{Mariia Gumberidze}
\affiliation{Department of Optics, Palacký University, 17. listopadu 12, 771 46, Olomouc, Czech Republic}

\author{Vladyslav Usenko}
\affiliation{Department of Optics, Palacký University, 17. listopadu 12, 771 46, Olomouc, Czech Republic}   
\author{Sven~Höfling}
    \affiliation{Technische Physik, University of Würzburg, Am Hubland, D-97074 Würzburg, Germany}

\author{Tobias~Huber-Loyola}
    \affiliation{Technische Physik, University of Würzburg, Am Hubland, D-97074 Würzburg, Germany}
    \affiliation{Institute of Photonics and Quantum Electronics, Karlsruhe Institute of Technology, Engesserstr. 5, 76131 Karlsruhe, Germany}
    \affiliation{Center for Integrated Quantum Science and Technology (IQST), Karlsruhe Institute of Technology, Karlsruhe, Germany}
\author{Mario~A.~Usuga~Castaneda}
    \affiliation{Single Quantum B.V., Delft, HH 2629, The Netherlands}

\author{Armando~Rastelli}
    \affiliation{Institute of Semiconductor and Solid State Physics, Johannes Kepler University, Altenbergerstraße 69, Linz 4040, Austria}

\author{Klaus~D.~Jöns}
    \affiliation{Institute for Photonic Quantum Systems (PhoQS), Center for Optoelectronics and Photonics Paderborn (CeOPP) and Department of Physics, Paderborn University, Warburger Straße 100, 33098, Paderborn, Germany}

\author{Rinaldo~Trotta}
\email{rinaldo.trotta@uniroma1.it}
    \affiliation{Dipartimento di Fisica, Sapienza Università di Roma, Piazzale Aldo Moro 5, 00185 Roma, Italy}
\date{}    

\begin{abstract}
Quantum key distribution (QKD) is a fundamental functionality of quantum networks. Extending its range requires quantum repeaters, a technology still under development and not yet demonstrated in combination with quantum cryptographic protocols. Here, we address the elementary operation of a viable quantum repeater scheme by evaluating whether correlated photons distributed via entanglement swapping can support secret key distribution. We employ a state-of-the-art deterministic photon source based on a strain-tuned GaAs quantum dot embedded in a photonic cavity and implement a source-independent QKD protocol between two nodes using entanglement-swapped photons. We benchmark the system and present how high-resolution temporal post-selection can enable secure key exchange, with further rate improvements expected through realistic device and protocol optimizations. Due to its compatibility with a wide range of quantum memory platforms, our approach represents a significant step towards long-distance quantum-safe communication.

\end{abstract}

\maketitle

\section{Introduction}
Quantum communication stands as one of the most impactful areas of quantum technologies, offering fundamentally new capabilities for information security and networked systems \cite{kimble2008quantum,wehner2018quantum}. Among its various applications, quantum key distribution (QKD) has emerged as the most mature and widely implemented one, enabling secure communication with information-theoretic guarantees \cite{pirandola2020advances, xu2020secure}. However, current QKD implementations face a significant limitation: the distance over which keys can be distributed is fundamentally constrained by photon losses in optical fibers or the atmosphere and the absence of signal amplification. As a result, long-distance QKD requires either trusted intermediate nodes or the development of quantum repeaters to preserve end-to-end security \cite{azuma2023quantum,briegel1998quantum,sangouard2011quantum,munro2015inside}. By dividing a long communication channel into shorter segments and using entanglement swapping to connect them, quantum repeaters promise to extend the reach of quantum communication without compromising its security. Despite substantial theoretical progress and encouraging experimental advances in individual components, quantum repeaters are limited to proof-of-concept experiments \cite{lagorivera2021telecom,liu2021heralded}, and crucially, no implementation to date has demonstrated support for a QKD protocol.
Several studies have explored this scenario through numerical simulations using photon pairs generated in nonlinear crystals \cite{khalique2015practical, guha2015rate},  both as a stepping stone towards a quantum repeater implementation and a way to mitigate the impact of detector dark noise \cite{scherer2011long}. Besides, heralded entanglement distribution granted by entanglement swapping is key to reach unconditional levels of security in QKD \cite{zapatero2023advances}, as it has been recently demonstrated employing entanglement swapping between spin qubits mediated by photons to enable device-independent secure QKD \cite{zhang2022device,nadlinger2022experimental,lu2026device,liu2026long}. However, an implementation based on all-photonic entanglement swapping is crucial to foster the realization of efficient chained protocols. This type of application has been recently demonstrated on a platform based on probabilistic sources \cite{jiang2026entanglement}, while it is still missing for quantum emitters, which would bring a potential advantage in device-independent heralded protocols \cite{zapatero2023advances, gisin2010proposal,kolodynski2020device}.\newline
\indent In this work, we experimentally address this challenge by presenting the successful combination of quantum emitter-based all-photonic entanglement swapping with an entanglement-based QKD protocol, which naturally integrates with the architecture of future entanglement-distribution networks. As the entangled photon source, we employ semiconductor quantum dots (QDs), which offer quasi-deterministic generation of entangled photon pairs \cite{jayakumar2013deterministic} and are compatible with many quantum memory platforms \cite{neuwirth2021quantum}, making them attractive candidates for solid-state quantum network nodes \cite{lodahl2017quantum,schimpf2021quantum,lu2021quantum,yu2023telecom}. In particular, GaAs QDs have been shown to provide high-rate, highly entangled photon pairs \cite{liu2019solid, wang2019demand, rota2024source}  at wavelengths that match well the working band of several quantum memory devices based on Rb atoms and atomic vapors \cite{chen2013coherent, namazi2017ultralow,wolters2017simple} or rare earths \cite{tang2015storage,davidson2020improved}, are compatible for free space and satellite mediated communication \cite{bourgoin_comprehensive_2013}, and can be used themselves as stationary qubits \cite{appel2025amany}.
We use entangled photon pairs from a state-of-the-art GaAs quantum dot embedded in a strain-tunable photonic-cavity structure \cite{rota2024source} to perform entanglement swapping with high fidelity, allowing us to employ the swapped photons to validate entanglement-based secure QKD. This result is achieved without resorting to entanglement distillation, but only performing temporal post-selection, which directly excludes the events resulting in unsuccessful outcomes of the swapping protocol, without acting on the photons distributed to the two communicating parties. 
This work represents a step forward in the integration of semiconductor QDs in quantum repeater architectures, providing a novel benchmark and motivating further research in the development of these emitters.

\section{Experimental results}

\begin{figure}[t]
    \centering
    \includegraphics[width=\linewidth]{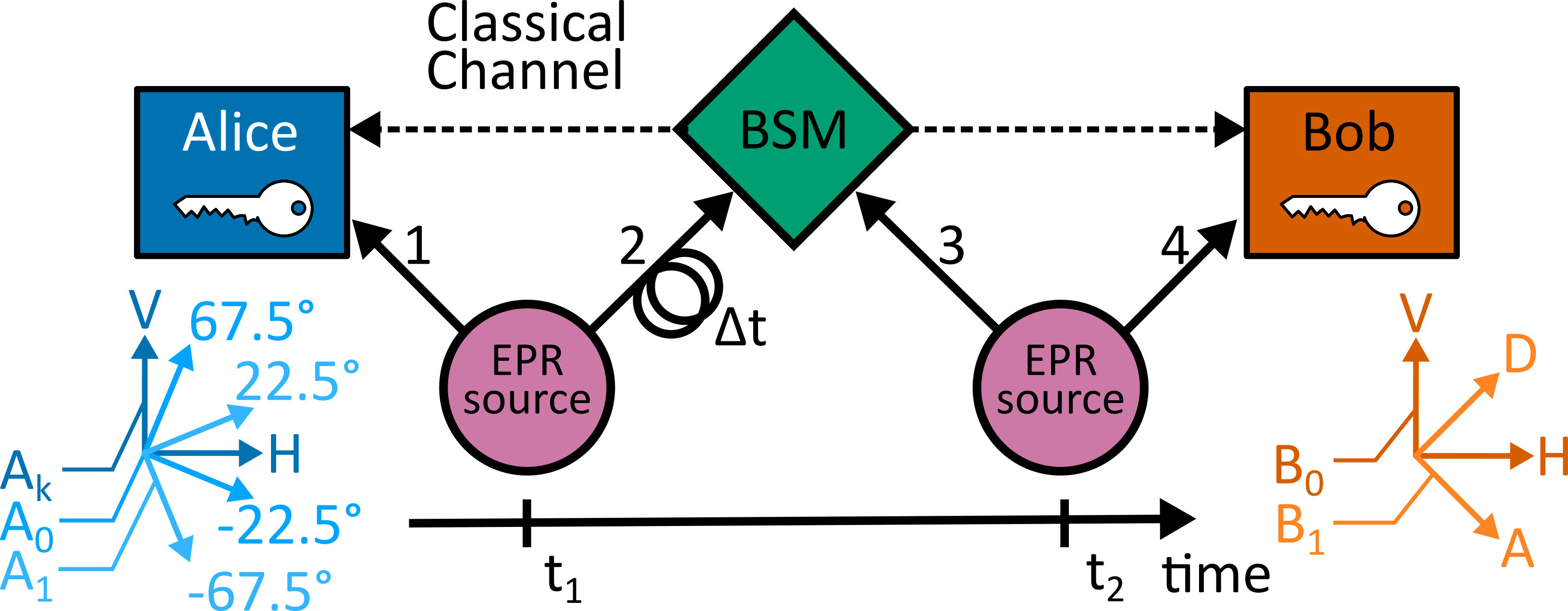}
    \caption{\textbf{Protocol scheme for quantum key distribution with swapped photons using a single entangled photon pair source.} \textit{A single EPR source is excited at two consecutive times $t_1$ and $t_2$, emitting each time a pair of entangled photons, labeled (1,2) and (3,4). Two of these entangled photons (2 and 3)  interfere in a 50\% efficiency Bell state measurement (BSM) setup by delaying the earlier photon via a delay line. The result of the BSM is shared among Alice and Bob with the pair of now mutually entangled photons 1 and 4. Using the information of the BSM, Alice and Bob perform measurements using different sets of bases, $A_k, A_0, \text{ and } A_1 $ for Alice and $B_0 \text{ and } B_1$ for Bob. Sharing the choice of measurement set and outcome of the BSM afterward allows for key sharing and checks on the Bell inequality violation.}}
    \label{fig:scheme}
\end{figure}

A scheme of the protocol is illustrated in Fig. \ref{fig:scheme}. The all-photonic implementation of entanglement swapping, akin to its first experimental demonstration \cite{pan1998experimental}, starts with two photon pairs (1,2 and 3,4) generated by the same entangled-photon pair source (EPR source) at subsequent times $t_1$ and $t_2$, each in the same maximally entangled Bell state, e.g., $\ket{\phi^+}=\frac{\ket{00}+\ket{11}}{\sqrt{2}}$. One photon is taken from each photon pair (2 and 3), and, after delaying the early photon by the difference in generation time $t_1-t_2$, they are subject to a projective Bell state measurement (BSM). As a result, the remaining two photons (1 and 4), which were previously uncorrelated, will be projected in a specific Bell state, which depends on the outcome of the BSM. This photon pair is shared between two communicating parties, together with the result of the BSM, which is sent through a classical information channel. In this way, the parties know which maximally entangled states they share, and they can exploit them as a resource to implement entanglement-based QKD. Here, we consider an asymmetrical variation of the Ekert protocol proposed by A. Ac\'{i}n and colleagues \cite{acin2006efficient, acin2007device}, that ensures source-independent security \cite{ekert1991quantum,pironio2009device}.
The two parties, denoted as Alice and Bob, each perform a single-qubit measurement, randomly choosing from an agreed set of bases, three for Alice and two for Bob. Relative to a common reference system of horizontal (H) and vertical (V) polarization corresponding to $0^\circ$ and $90^\circ$ polarization angle, the measurement bases are \{$A_k, A_0, A_1$\} = \{$0^\circ/90^\circ$, $-22.5^\circ/67.5^\circ$, $-67.5^\circ/22.5^\circ$\} and \{$B_0, B_1$\}  = \{$0^\circ/90^\circ$, $-45^\circ/45^\circ$\}\cite{ling2008experimental}. The aligned pairing of bases \{$A_k, B_0$\} is used to reconstruct the shared key. The combinations of the remaining measurement channels \{$A_0/A_1, B_0/B_1$\} are used to verify the Clauser–Horne-Shimony–Holt (CHSH) violation \cite{clauser1969proposed} and quantify possible eavesdropping.\\
\indent The protocol performance can be affected by non-ideal preparation of the shared state and measurement operations. In general, these kinds of flaws cannot be distinguished from the effects of the malicious action of an eavesdropper; for this reason, any loss of mutual information between Alice and Bob regarding the key due to experimental imperfections must be assimilated to a gain in information for an adversary party, denoted as Eve. The state preparation process is, in this case, the result of an entanglement swapping protocol: the accuracy of the operation, quantified by the fidelity of the obtained state to the expected one, mainly depends on the characteristics of the photon source \cite{basso2019entanglement, zopf2019entanglement}.\\
\indent The emitter we employ is a QD device that can generate strongly polarization-entangled photon pairs in a near-deterministic fashion via the biexciton-exciton cascade under resonant two-photon excitation \cite{jayakumar2013deterministic, muller2014ondemand}. In the following, we denote the biexciton to exciton transition as XX and the exciton to ground-state transition as X. As the material platform, we choose GaAs/AlGaAs QDs fabricated by droplet etching epitaxy, which have been proven to generate strongly entangled photon pairs \cite{huber2018strain}, arguably due to their short lifetimes, low strain gradient, and moderate spin noise \cite{schimpf2023hyperfine}. Additionally, we use a state-of-the-art device in which the QD is embedded in a circular Bragg resonator (CBR) to enhance brightness and is integrated on a micromachined piezoelectric substrate to enable fine structure splitting (FSS) removal \cite{rota2024source}, which is an important requirement to observe highly entangled photons. The emission is characterized by a low multi-pair component \cite{hanschke2018quantum, neuwirth2022multipair}, granting an advantage in quantum cryptographic applications, as compared to probabilistic approaches based on spontaneous parametric down-conversion (SPDC) light sources \cite{basso2021quantum, hosak2021effect,gumberidze2025feasibilityentanglementbasedqkdprotocols}. 

For the emitter used in this experiment (see Ref. \cite{laneve2024quantum} for a full characterization of the QD) we measure a FSS$=\SI{0.3\pm0.2}{\micro\electronvolt}$ via polarization-resolved photoluminescence, which, combined with a Purcell-enhanced X transition lifetime $\tau_X=\SI{100(10)}{\ps}$, results in the emission of photon pairs with a \num{0.94 \pm 0.01} fidelity to the $\ket{\phi^+}$ state, as acquired with a full quantum state tomography \cite{altepeter2005photonic}. 
In addition to fidelity to a maximally entangled state, the performance of an entanglement swapping protocol depends on photon indistinguishability, since the efficient optical implementation of polarization-encoded BSM relies on successful two-photon interference. High indistinguishability is more challenging to achieve, especially for photons generated via the XX-X cascade that are linked by energy-time entanglement \cite{simon2005creating,scholl2020crux}. We estimate indistinguishability by measuring the interference visibility in a Hong-Ou-Mandel (HOM) experiment using photons from the exciton to ground-state transition excited consecutively with a \SI{1.6}{\nano\second} delay. The HOM visibility we witness is limited to  \num{0.520\pm0.002}, due to a X transition linewidth $\Delta\text{E} =\SI{22(2)}{\micro\eV}$ (more than 6 times its Fourier limit) and which-path information inherent to two-photon cascaded emission \cite{scholl2020crux}, given by the ratio between X and XX lifetimes ($\tau_{XX}=\SI{67(5)}{\ps}$). 
As already demonstrated in past literature \cite{legero2004quantum}, HOM visibility can be significantly increased by using low-time jitter detectors and employing temporal post-selection at the cost of retaining a fraction of the total photon detection events. This technique was used to achieve quantum interference between QD photons  \cite{patel2008postselective,patel2010two}, and then exploited in quantum teleportation experiments involving QD light \cite{huwer2017quantum,anderson2020quantum, anderson2020gigahertz}, recently interfacing remote sources \cite{laneve2024quantum, strobel2024quantum} and entanglement swapping with single \cite{basso2019entanglement, zopf2019entanglement} and remote sources \cite{beccaceci_all-photonic_2025}. 
By selecting a narrower time window for the coincidence events of our HOM experiment, the measured visibility can be increased up to \num{0.79\pm0.02} (see Supplemental Material for the complete study of the dependence on the coincidence window).
Relying on theoretical modeling, we can expect a source with such features to be able to support a significant violation of Bell's inequality using swapped photons \cite{rota2020entanglement,schimpf2021quantum}.\\
\indent Before proceeding to QKD validation, we assess the fidelity of state preparation resulting from entanglement swapping with photons from our QD. We generate two sequential entangled photon pairs, exciting the same QD with two laser pulses separated by a delay of \SI{1.6}{\nano\second}. This double excitation is repeated at a rate of \SI{160}{\MHz}. X photons are separated using a volume Bragg grating filter and sent to an unbalanced Mach-Zehnder interferometer, with the same excitation time delay on one of the arms, to induce interference between X photons emitted by distinct cascades, which we denote as early (X$_e$) and late (X$_l$) X photons. The BSM setup is composed of a balanced BS and two polarizing beam splitters (PBSs), which allow the discrimination of two Bell states, that is the optimal strategy which can be implemented using only linear optics \cite{lutkenhaus1999bell}.
\begin{figure}[t]
    \centering
\includegraphics[width=\linewidth]{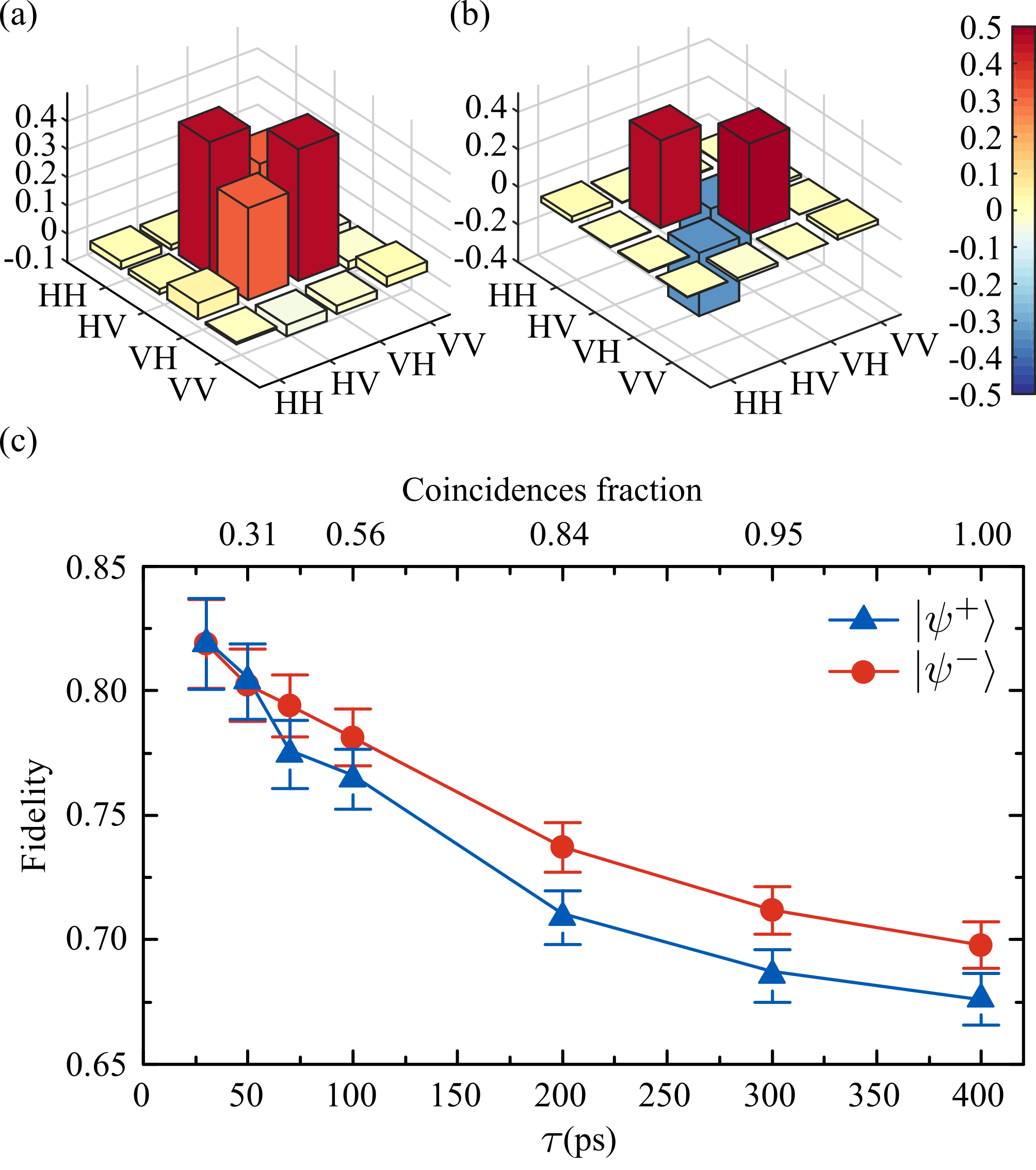}
    \caption{\textbf{Experimental entanglement swapping performance.} \textit{(a) Real part of the density matrix of the biexciton photons given a $\ket{\psi^+}$ BSM measured with a 30ps time window on the BSM. (b) Same as in (a) for a $\ket{\psi^-}$ BSM. (c) Fidelity of the biexciton photons state to the expected state as a function of the coincidence time window applied to the BSM events. The retained fraction of coincidence counts for different time windows is reported on the top axis. The total 4-fold coincidences corresponding to the largest coincidence window, i.e., \SI{400}{\ps}, are \SI{22958}, corresponding to a coincidence rate of \SI{6.4}{\per\second}.}}
\label{fig:entanglement_swap}
\end{figure}
Conditioned on the detection of one of the two outcomes of the BSM, a full quantum state tomography is performed on the remaining XX photon pairs - composed of the early (XX$_e$) and late (XX$_l$) XX photons - which are randomly separated with a 50:50 BS, to reconstruct their two-qubit polarization state. We analyze the fidelity of the resulting state to the expected Bell state as a function of the BSM coincidence temporal window. The results are reported in Fig. \ref{fig:entanglement_swap}. As expected, the fidelity obtained with a large coincidence window is limited to $\num{0.69 \pm 0.01}$, which is arguably insufficient to support an entanglement-based QKD protocol \cite{ekert1991quantum, ma2007quantum,davis2025entanglement}. However, the fidelity improves with narrower BSM coincidence windows, at the cost of a lower coincidence rate. These results are compatible with a previous similar study \cite{zopf2019entanglement} with the additional advantage of demonstrating the same operation in a state-of-the-art device with embedding in an optical microcavity for enhanced brightness and strain-tuning capabilities.
\begin{figure}[t]
    \centering
    \includegraphics[width=\linewidth]{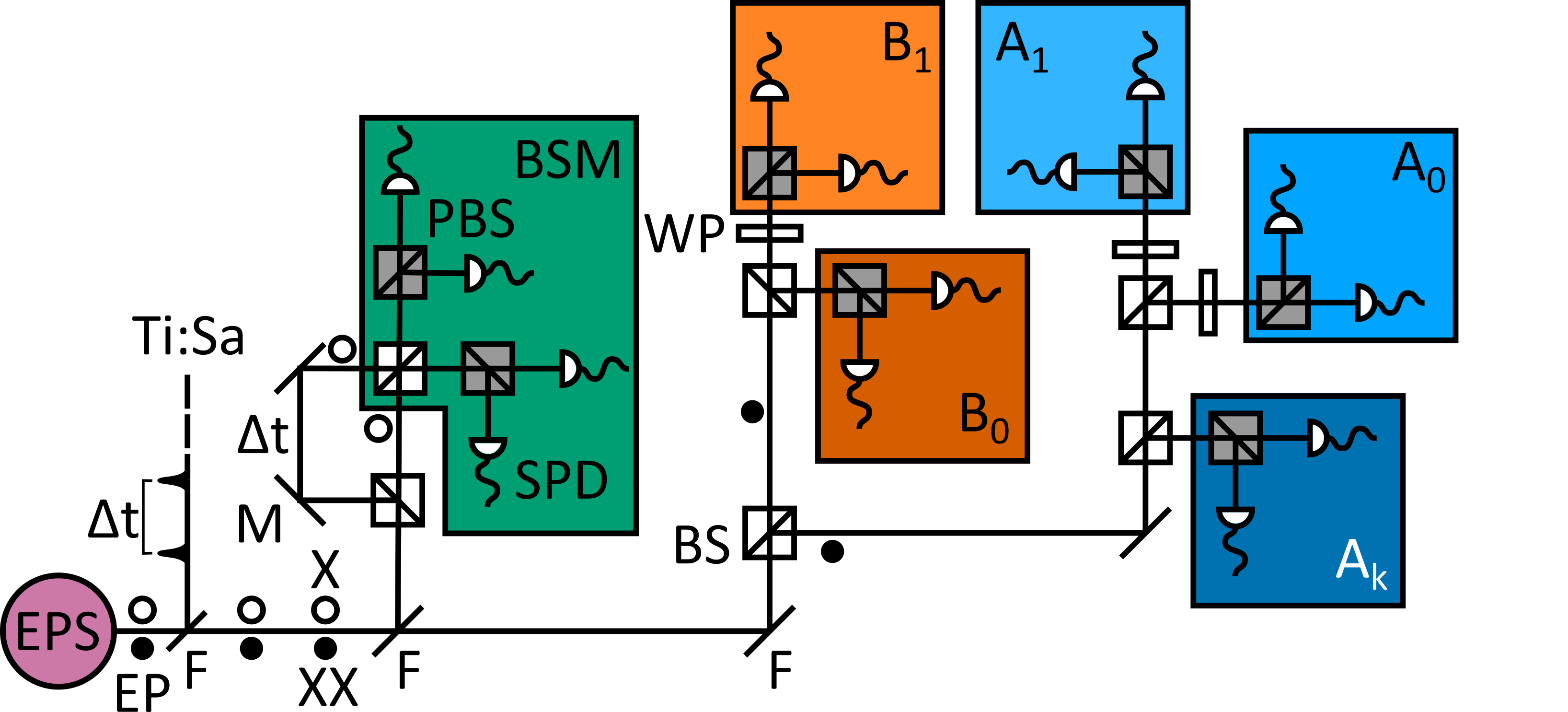}
    \caption{\textbf{Scheme of the QKD experimental setup}. \textit{The entangled pair source (EPS) is excited by a pulsed Ti:Sapphire laser (Ti:Sa), doubled in repetition rate by an unbalanced Mach-Zehnder interferometer with a time delay $\Delta t$. The emitted pairs of photons, hollow (X photons) and full (XX photons) circles, are separated employing spectral filters (F) and sent respectively to the Bell state measurement (BSM) setup and Alice and Bob. In the BSM, the state of the photons from subsequent pulses is projected into two possible Bell states by making them interfere in another unbalanced Mach-Zehnder interferometer with the same $\Delta t$ and measuring coincidence combinations at 4 single photon detectors (SPDs) placed at the exits of two polarizing BSs (PBS). The remaining photons, in an entangled state after the BSM, are randomly sent to Alice and Bob setups, where their polarization state is measured in different, randomly chosen polarization bases; $\{A_0, A_1, A_k\}$ and $\{B_0, B_1\}$ by using SPDs, PBSs and half-waveplates (WP).}}
    \label{fig:setup}
\end{figure}

\begin{figure}[t]
    \centering
    \includegraphics[width=\linewidth]{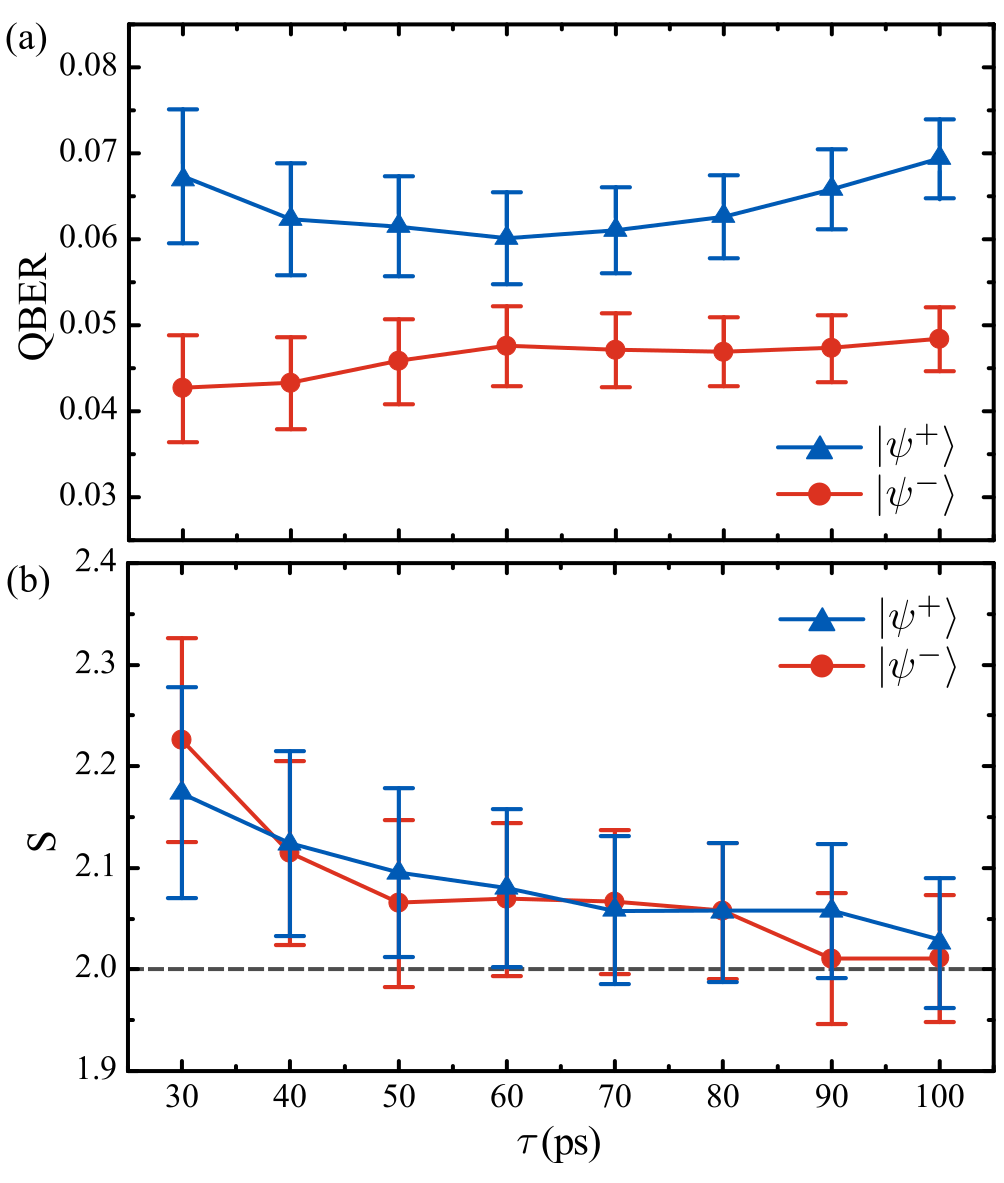}
    \caption{\textbf{QKD performance as a function of BSM coincidence time window.} (a) QBER as a function of the BSM coincidence time window. (b) Bell parameter $S$ as a function of the BSM coincidence time window. The threshold $S=2$ for classical correlation is highlighted by a dashed grey line.} 
    \label{fig:qkd_vs_timewindow}
\end{figure}

To perform entanglement-based QKD, XX photons are randomly separated by the BS and distributed instead to Alice and Bob, so that the two parties will probabilistically share an XX$_e$ and XX$_l$ photon pair.
The experimental setup employed to realize the full QKD experiment with entanglement-swapped photons is schematically reported in Fig. \ref{fig:setup}.
Alice and Bob implement the protocol measurement bases through BSs (to randomize the measurement choice) and combinations of half-wave plates (HWPs) and PBSs to perform the selected polarization state projections. 8 single-nanowire single-photon detectors (SNSPDs) from Single Quantum with $<\SI{15}{\ps}$ time resolution are used for the 4 channels of the BSM and the 4 channels for key-sharing, i.e., $\{B_0, A_k\}$. 6 single photon avalanche photodiodes (SPADs) with $<\SI{350}{\ps}$ time resolution are instead used for the remaining channels used for the check of Bell inequality violation, i.e., $\{B_1, A_1, A_0\}$. The signal from all the detectors is sent to a time-to-digital converter (Time Tagger Ultra from Swabian Instruments) with a nominal \SI{19}{\ps} time resolution per channel. Only time tags related to fourfold coincidence events within a \SI{15}{\ns} window are retained to reduce the size of saved data and speed up post-measurement analysis. To evaluate the performance of the QKD protocol, we monitor two crucial parameters from the relevant combinations of BSM-heralded two-photon correlations, namely the quantum bit error rate (QBER) and the Bell parameter ($S$). The QBER quantifies the discrepancy between the two copies of the raw key after sifting and is defined as $Q = (1 - E(A_k, B_0))/2$, where $E$ is the expectation value calculated from the number of coincidences measured on the different combinations of outcomes of the two given measuring bases. The parameter $S$ quantifies the violation of the computed CHSH inequality \cite{clauser1969proposed,acin2007device,pironio2009device}, defined, for our bases choice and distributed states, as: $S^{\psi^\pm}=-E(A_0, B_0)\pm E(A_0, B_1) + E(A_1, B_0) \pm E(A_1, B_1)$. The parameters have been studied again as a function of the BSM coincidence window to identify the optimal choice to maximize QKD efficiency.
We can see in Fig. \ref{fig:qkd_vs_timewindow}(a) that the QBER is relatively low, within the requirements for secure entanglement-based QKD \cite{ekert1991quantum, acin2007device}, and is almost unaffected by the BSM time window. This fact stems from the employment of a 50\% efficiency, polarization-selective BSM, and the specific choice of Alice and Bob to align the key channels, i.e., \{$A_k, B_0$\}, to the same polarization axes of the BSM ($H, V$). It can be shown \cite{basso2019entanglement, basso2021quantum} that, in this particular case, the expected values yielding the QBER (corresponding to the operators $\ket{HV}\bra{HV}$ and $\ket{VH}\bra{VH}$) are theoretically independent from the indistinguishability of the two swapped photons, when computed on their density matrix (explicitly reported in the SM). In these circumstances, the QBER coefficient is minimized because it depends almost exclusively on the degree of entanglement of the photon pairs rather than on their indistinguishability, the former being closer to unity for our photon source.
In turn, correlations will generally be weaker when measured in other non-commuting bases due to the quantum state discrimination performed by the BSM, which also relies on indistinguishability. Therefore, the Bell parameter $S$ will be affected by the BSM time window. The recorded experimental values of the Bell parameter are reported in Fig. \ref{fig:qkd_vs_timewindow}(b). For the largest BSM coincidence window, the Bell parameter is lower than the threshold for classical correlations, but it is projected to exceed it for narrower time windows.
Given these figures of merit, we now evaluate how the performed data exchange can support sharing a secret key in the asymptotic limit. In particular, we use the Devetak-Winter (DW) rate as a lower bound to estimate the fraction of key rate that can be safely considered secure. Under the common assumption of Alice and Bob controlling their measuring apparatuses, this quantity is given by $r_{DW}=1-h(Q)-h(Q+S/(2\sqrt{2}))$, where $h$ is the binary Shannon entropy \cite{acin2007device}. This formula does not include the efficiency of the error correction algorithm, which depends on the dataset and the specific chosen implementation \cite{mateo2015demystifying}. We define then the secure fraction of the distributed key (SKF) as SKF$=\max(0,r_{DW})$, since no part of the key can be assumed to be safe from an eavesdropper if $r_{DW}<0$. The value of SKF is calculated as a function of the BSM coincidence window. As expected from the results of the violation of Bell's inequality, it is impossible to support a secure key exchange for large BSM coincidence time windows. It is often necessary to exceed a simple violation, depending on the specific value of QBER. The condition is satisfied for a \SI{40}{\ps} or narrower window, for the $\ket{\psi^-}$ state only, and we experimentally obtain the maximum positive asymptotic SKF of 9(8)\% for a \SI{30}{\ps} window.
This result demonstrates that the current figures of merit of QD-based photon sources could already support secure communication in a scenario where entanglement is further distributed using entanglement swapping protocols.\\
\indent In a quantum networking framework, it is relevant to consider the combination of our approach with entanglement distillation \cite{deutsch1996quantum,dur1999quantum}, a key element in quantum repeater architectures \cite{bratzik2013quantum}. This operation usually requires efficient quantum memories and reliable application of two-qubit gates, so that an approach based on improving the source features, as ours, might deliver an efficiency improvement both in combination with entanglement distillation procedures and as an alternative to them. Simulating the application of a basic entanglement distillation protocol \cite{deutsch1996quantum} to the experimental density matrices obtained as the outcome of entanglement swapping, we estimate that, by applying distillation on top of our time-wise filtering procedure, it would be possible to achieve a total SKF of up to $15.7\%$ for a coincidence time window of $30$ ps, including in this value the success rate of the distillation protocol and a $\frac{1}{2}$ factor due to the requirement of expending two photon pairs for each protocol run, to get as an outcome a single (more entangled) photon pair. This value corresponds to an average estimated fidelity (considering both $\ket{\psi^+}$ and $\ket{\psi^-}$ projections) of $0.92$. 
In the SM, we report this analysis in detail and show how the distilled entanglement changes as a function of the time window applied to the BSM, for both $\ket{\psi^+}$ and $\ket{\psi^-}$ projections.
We note that this distillation analysis is performed on the experimentally reconstructed density matrices assuming ideal local operations, and its practical implementation would involve experimentally significant requirements.\\
\section{Discussion}
Our results provide several valuable insights into which steps forward are required to make the presented approach practical. The first consideration is related to the overall secret key rate (SKR), defined as the product between the sifted key rate $R_K$ and the SKF, SKR$=R_K*$SKF. The sifted key rate depends on fourfold coincidence events, which are strongly affected by source inefficiency and setup losses. The optimal SKR is given by a tradeoff between the sifted key rate, which strictly decreases for narrower BSM coincidence windows, and the SKF, which improves for smaller windows. The optimal time window is obtained for a value of \SI{30}{\ps} corresponding to an estimated secure key rate of \SI{1E-3}{\bit/\second}. We want to underline that the chosen temporal window, which maximizes the SKR, does not maximize the photon indistinguishability. We point out that this rate makes the sharing of long keys still impractical, which is the reason why we focused our considerations on the asymptotic limit. 
It is interesting to consider how additional processing of the swapped photons might change the key rate. For instance
 applying a simple distillation protocol, which is often included in quantum repeater architectures \cite{azuma2023quantum, goodenough2021optimizing, mantri2025comparing}, and even for QKD applications \cite{bratzik2013quantum}, we estimate that it would be possible to more than double the secure key rate, mainly due to the improvement in the $r_{DW}$ available for the projection $\ket{\psi^+}$. However, in this case, the optimal coincidence window would be $40$ps, due to the inevitable trade-off between the losses and the swapping fidelity improvements induced by narrower BSM time windows.
Regarding the effective security of such an obtained key, it is important to discuss the specific attacks from an eavesdropper that the time-wise filtering we perform might enable. We can consider the case in which such a malicious party had access to the events that were discarded by the communicating parties, namely, possible key events triggered by BSM coincidences featuring a delay larger than the selected window. We consider Alice to be the reference party for the sifted key. We consider then the key string that she would obtain if she considered a 30ps time window on the BSM or a 400ps one (that comprises almost all the four-fold coincidences). From the latter, we subtract the events which correspond to the 30ps key, because Alice will keep only them, while the rest we suppose to be available to an eavesdropper, Eve. By computing the mutual information between the keys now in the hands of Alice and Eve, we find out that Eve can extract a negligible amount of information from her string to guess Alice's one (details are provided in the SM).
On the other hand, some improvements are well within reach to significantly increase the key rate.\\
\indent The largest opportunity for improvement is related to the brightness of the source. More specifically, the performance of the QD used in this experiment can be mainly improved in three aspects: blinking, extraction efficiency, and coupling efficiency into a single-mode fiber. Even limiting ourselves to combining state-of-the-art figures achieved for broadband QD sources, that is, negligible blinking \cite{schimpf2021entanglement}, 85\% extraction efficiency \cite{liu2019solid}, and 65\% single-mode coupling efficiency \cite{wang2019demand}, would increase the single-photon rates by an order of magnitude and four-fold coincidences by about a factor of 350. Also, the pumping rate can be increased from \SI{320}{\MHz} to the GHz range due to the fast sub-ns emission times of GaAs QDs.\\
\indent A further significant area for improvement is linked to the fidelity of the state obtained from the entanglement swapping operation. Since the experimental secret key fraction is very close to the success threshold of the protocol, even small gains in QBER and Bell violation would lead to a much improved key throughput. As already mentioned, at a source level, the main challenge is related to photon indistinguishability. A QD with better optical quality should already provide a performance closer to the limit imposed by the presence of time correlations in the radiative cascade \cite{scholl2020crux}. Current research is aiming to improve this limit through photonic engineering of the transition lifetimes, by accelerating the first step of the cascade \cite{huber2013measurement}; the application of external electric fields has been recently proved remarkably effective to this purpose \cite{undeutsch2025electric}.
A better tradeoff in temporal filtering could be achieved as well with the use of a cutting-edge time-to-digital converter that can deliver a lower time resolution compared to the \SI{19}{\ps} (FWHM) of our current device. 
More practical scenarios will require the use of separated emitters, avoiding the probabilistic separation of the photon streams - a necessary part of our experimental design - but increasing the technical hurdles related to indistinguishability. However, the challenge of remote two-photon interference has been widely explored for QD sources, leading to the recent demonstrations of single-emitter-like performance in electrically controlled devices with high spectral stability \cite{zhai2022quantum} and of application in a benchmarking protocol such as quantum teleportation, even in practical scenarios \cite{laneve2024quantum, strobel2024quantum}. 
Future devices able to combine electrical control with high-rate generation of entangled photon pairs will be able to successfully tackle integration into realistic network architectures.
Although free-space networks remain the most suitable application framework for the emitters we employ \cite{laneve2024quantum}, thanks to laser-mediated frequency conversion it is possible to match the optimal wavelength for fiber transmission \cite{ikuta2018polarization,weber2019two,strobel2024quantum}.
High fidelity of the distributed entangled state is important to ensure that a QKD protocol can be sustained in a complete quantum repeater. Enough performance headroom needs to be available to account for the below-unity fidelity of the quantum memories used to store the qubits while required classical information is exchanged \cite{neuwirth2021quantum}, although promising broadband quantum memories with fidelity above 95\% are already under development \cite{davidson2020improved}.\\
\indent Further insights could be gained by investigating relevant security aspects, such as quantitative analysis of possible side channels \cite{xu2020secure} and comparative studies of the viable choices for key refinement algorithms \cite{tomamichel2017largely,lim2021security}.\\
\indent In conclusion, our work is a signifcant step towards the reliable implementation of QKD protocols over long distances, enabled by quantum repeaters.  Quantum dot technology can play an important role in this outlook in combination with compatible high-performance hardware for quantum information storage and detection. This system integration and the anticipated advances in photon generation devices will lead to the practical implementation of quantum cryptography protocols into the next generation of repeater-assisted communication networks.
\newline

M.B.R. and F.B.B. contributed equally to this work.

\acknowledgments The authors thank Dr. Edoardo Mornacchi from Swabian Instruments and A. Miriametro for the technical support  and K. Broccoletti for administrative support. This project has received funding from the European Union’s Horizon 2020 Research and Innovation Program under Grant Agreement n. 899814 (Qurope) and from the QuantERA II Programme that has received funding from the European Union’s Horizon 2020 research and innovation programme under Grant Agreement n. 101017733 via the project QD-E-QKD. The authors also acknowledge support from MUR (Ministero dell’Università e della Ricerca) through the PNRR MUR project PE0000023-NQSTI, the European Union’s Horizon Europe research and innovation program under the EPIQUE Project GA No. 101135288. S.F.C. da Silva acknowledges São Paulo Research Foundation (FAPESP), Brasil, Process Number 2024/08527-2 and 2024/21615-8 for financial support. AR acknowledges support of the EU HE EIC Pathfinder challenges action under grant agreement No. 101115575 (Q-ONE), from the QuantERA program via the project MEEDGARD (FFG Grant No. 906046) the Austrian Science Fund FWF via the Research Group FG5 [10.55776/FG5] and the cluster of excellence Quantum Science Austria (quantA) [10.55776/COE1] as well as the Linz Institute of Technology (LIT), and the LIT Secure and Correct Systems Lab, supported by the State of Upper Austria. MG and VU acknowledge support from the European Union’s Horizon Europe research and innovation programme under the project QSNP, GA No. 101114043, and support from the Czech Science Foundation under the project No. 21-44815L. This work is supported by the German Federal Ministry of Research, Technology and Space (BMFTR) through the project QR.N (16KIS2206). This project has received funding from the European Research Council (ERC) under the European Union’s Horizon 2020 researchand innovation program (LiNQs, grant agreement 101042672). T.H.L. acknowledges funding from the German ministry of research, technology and space (BMFTR) for the project Qecs (FKZ: 13N16272).

\bibliography{main}

@article{appel2025amany,
  title = {A many-body quantum register for a spin qubit},
  volume = {21},
  ISSN = {1745-2481},
  url = {http://dx.doi.org/10.1038/s41567-024-02746-z},
  DOI = {10.1038/s41567-024-02746-z},
  number = {3},
  journal = {Nature Physics},
  publisher = {Springer Science and Business Media LLC},
  author = {Appel,  Martin Hayhurst and Ghorbal,  Alexander and Shofer,  Noah and Zaporski,  Leon and Manna,  Santanu and da Silva,  Saimon Filipe Covre and Haeusler,  Urs and Le Gall,  Claire and Rastelli,  Armando and Gangloff,  Dorian A. and Atat\"{u}re,  Mete},
  year = {2025},
  month = jan,
  pages = {368–373}
}

@article{pirandola2020advances,
  title = {Advances in quantum cryptography},
  volume = {12},
  ISSN = {1943-8206},
  url = {http://dx.doi.org/10.1364/AOP.361502},
  DOI = {10.1364/aop.361502},
  number = {4},
  journal = {Advances in Optics and Photonics},
  publisher = {Optica Publishing Group},
  author = {Pirandola,  S. and Andersen,  U. L. and Banchi,  L. and Berta,  M. and Bunandar,  D. and Colbeck,  R. and Englund,  D. and Gehring,  T. and Lupo,  C. and Ottaviani,  C. and Pereira,  J. L. and Razavi,  M. and Shamsul Shaari,  J. and Tomamichel,  M. and Usenko,  V. C. and Vallone,  G. and Villoresi,  P. and Wallden,  P.},
  year = {2020},
  month = dec,
  pages = {1012}
}

@article{lagorivera2021telecom,
  title = {Telecom-heralded entanglement between multimode solid-state quantum memories},
  volume = {594},
  ISSN = {1476-4687},
  url = {http://dx.doi.org/10.1038/s41586-021-03481-8},
  DOI = {10.1038/s41586-021-03481-8},
  number = {7861},
  journal = {Nature},
  publisher = {Springer Science and Business Media LLC},
  author = {Lago-Rivera,  Dario and Grandi,  Samuele and Rakonjac,  Jelena V. and Seri,  Alessandro and de Riedmatten,  Hugues},
  year = {2021},
  month = jun,
  pages = {37–40}
}

@article{liu2021heralded,
  title = {Heralded entanglement distribution between two absorptive quantum memories},
  volume = {594},
  ISSN = {1476-4687},
  url = {http://dx.doi.org/10.1038/s41586-021-03505-3},
  DOI = {10.1038/s41586-021-03505-3},
  number = {7861},
  journal = {Nature},
  publisher = {Springer Science and Business Media LLC},
  author = {Liu,  Xiao and Hu,  Jun and Li,  Zong-Feng and Li,  Xue and Li,  Pei-Yun and Liang,  Peng-Jun and Zhou,  Zong-Quan and Li,  Chuan-Feng and Guo,  Guang-Can},
  year = {2021},
  month = jun,
  pages = {41–45}
}

@article{khalique2015practical,
  title = {Practical long-distance quantum key distribution through concatenated entanglement swapping with parametric down-conversion sources},
  volume = {32},
  ISSN = {1520-8540},
  url = {http://dx.doi.org/10.1364/JOSAB.32.002382},
  DOI = {10.1364/josab.32.002382},
  number = {11},
  journal = {Journal of the Optical Society of America B},
  publisher = {Optica Publishing Group},
  author = {Khalique,  Aeysha and Sanders,  Barry C.},
  year = {2015},
  month = oct,
  pages = {2382}
}

@article{guha2015rate,
  title = {Rate-loss analysis of an efficient quantum repeater architecture},
  volume = {92},
  ISSN = {1094-1622},
  url = {http://dx.doi.org/10.1103/PhysRevA.92.022357},
  DOI = {10.1103/physreva.92.022357},
  number = {2},
  journal = {Physical Review A},
  publisher = {American Physical Society (APS)},
  author = {Guha,  Saikat and Krovi,  Hari and Fuchs,  Christopher A. and Dutton,  Zachary and Slater,  Joshua A. and Simon,  Christoph and Tittel,  Wolfgang},
  year = {2015},
  month = aug 
}

@article{scherer2011long,
  title = {Long-distance practical quantum key distribution by entanglement swapping},
  volume = {19},
  ISSN = {1094-4087},
  url = {http://dx.doi.org/10.1364/OE.19.003004},
  DOI = {10.1364/oe.19.003004},
  number = {4},
  journal = {Optics Express},
  publisher = {Optica Publishing Group},
  author = {Scherer,  Artur and Sanders,  Barry C. and Tittel,  Wolfgang},
  year = {2011},
  month = feb,
  pages = {3004}
}

@article{ekert1991quantum,
  title={Quantum cryptography based on {B}ell’s theorem},
  author={Ekert, Artur K},
  journal={Physical review letters},
  volume={67},
  number={6},
  pages={661},
  year={1991},
  publisher={APS}
}

@article{kimble2008quantum,
  title={The quantum internet},
  author={Kimble, H Jeff},
  journal={Nature},
  volume={453},
  number={7198},
  pages={1023--1030},
  year={2008},
  publisher={Nature Publishing Group}
}

@article{wehner2018quantum,
  title={Quantum internet: A vision for the road ahead},
  author={Wehner, Stephanie and Elkouss, David and Hanson, Ronald},
  journal={Science},
  volume={362},
  number={6412},
  pages={eaam9288},
  year={2018},
  publisher={American Association for the Advancement of Science}
}

@article{munro2015inside,
  title={Inside quantum repeaters},
  author={Munro, William J and Azuma, Koji and Tamaki, Kiyoshi and Nemoto, Kae},
  journal={IEEE Journal of Selected topics in quantum electronics},
  volume={21},
  number={3},
  pages={78--90},
  year={2015},
  publisher={IEEE}
}

@article{lim2021security,
  title = {Security Analysis of Quantum Key Distribution with Small Block Length and Its Application to Quantum Space Communications},
  author = {Lim, Charles Ci-Wen and Xu, Feihu and Pan, Jian-Wei and Ekert, Artur},
  journal = {Phys. Rev. Lett.},
  volume = {126},
  issue = {10},
  pages = {100501},
  numpages = {5},
  year = {2021},
  month = {Mar},
  publisher = {American Physical Society},
  doi = {10.1103/PhysRevLett.126.100501},
  url = {https://link.aps.org/doi/10.1103/PhysRevLett.126.100501}
}

@article{tomamichel2017largely,
  title={A largely self-contained and complete security proof for quantum key distribution},
  author={Tomamichel, Marco and Leverrier, Anthony},
  journal={Quantum},
  volume={1},
  pages={14},
  year={2017},
  publisher={Verein zur F{\"o}rderung des Open Access Publizierens in den Quantenwissenschaften}
}

@article{xu2020secure,
  title = {Secure quantum key distribution with realistic devices},
  author = {Xu, Feihu and Ma, Xiongfeng and Zhang, Qiang and Lo, Hoi-Kwong and Pan, Jian-Wei},
  journal = {Rev. Mod. Phys.},
  volume = {92},
  issue = {2},
  pages = {025002},
  numpages = {60},
  year = {2020},
  month = {May},
  publisher = {American Physical Society},
  doi = {10.1103/RevModPhys.92.025002},
  url = {https://link.aps.org/doi/10.1103/RevModPhys.92.025002}
}

@article{wang2019demand,
  title={On-demand semiconductor source of entangled photons which simultaneously has high fidelity, efficiency, and indistinguishability},
  author={Wang, Hui and Hu, Hai and Chung, T-H and Qin, Jian and Yang, Xiaoxia and Li, J-P and Liu, R-Z and Zhong, H-S and He, Y-M and Ding, Xing and others},
  journal={Physical review letters},
  volume={122},
  number={11},
  pages={113602},
  year={2019},
  publisher={APS}
}

@article{sangouard2011quantum,
  title={Quantum repeaters based on atomic ensembles and linear optics},
  author={Sangouard, Nicolas and Simon, Christoph and De Riedmatten, Hugues and Gisin, Nicolas},
  journal={Reviews of Modern Physics},
  volume={83},
  number={1},
  pages={33--80},
  year={2011},
  publisher={APS}
}

@article{azuma2023quantum,
  title={Quantum repeaters: From quantum networks to the quantum internet},
  author={Azuma, Koji and Economou, Sophia E and Elkouss, David and Hilaire, Paul and Jiang, Liang and Lo, Hoi-Kwong and Tzitrin, Ilan},
  journal={Reviews of Modern Physics},
  volume={95},
  number={4},
  pages={045006},
  year={2023},
  publisher={APS}
}

@article{lodahl2017quantum,
  title={Quantum-dot based photonic quantum networks},
  author={Lodahl, Peter},
  journal={Quantum Science and Technology},
  volume={3},
  number={1},
  pages={013001},
  year={2017},
  publisher={IOP Publishing}
}

@article{neuwirth2021quantum,
  title={Quantum dot technology for quantum repeaters: from entangled photon generation toward the integration with quantum memories},
  author={Neuwirth, Julia and Basso Basset, Francesco  and Rota, Michele B and Roccia, Emanuele and Schimpf, Christian and J{\"o}ns, Klaus D and Rastelli, Armando and Trotta, Rinaldo},
  journal={Materials for Quantum Technology},
  volume={1},
  number={4},
  pages={043001},
  year={2021},
  publisher={IOP Publishing}
}

@article{davidson2020improved,
  title = {Improved light-matter interaction for storage of quantum states of light in a thulium-doped crystal cavity},
  author = {Davidson, Jacob H. and Lefebvre, Pascal and Zhang, Jun and Oblak, Daniel and Tittel, Wolfgang},
  journal = {Phys. Rev. A},
  volume = {101},
  issue = {4},
  pages = {042333},
  numpages = {5},
  year = {2020},
  month = {Apr},
  publisher = {American Physical Society},
  doi = {10.1103/PhysRevA.101.042333},
  url = {https://link.aps.org/doi/10.1103/PhysRevA.101.042333}
}

@article{yu2023telecom,
  title={Telecom-band quantum dot technologies for long-distance quantum networks},
  author={Yu, Ying and Liu, Shunfa and Lee, Chang-Min and Michler, Peter and Reitzenstein, Stephan and Srinivasan, Kartik and Waks, Edo and Liu, Jin},
  journal={Nature nanotechnology},
  volume={18},
  number={12},
  pages={1389--1400},
  year={2023},
  publisher={Nature Publishing Group UK London}
}

@article{liu2019solid,
  title={A solid-state source of strongly entangled photon pairs with high brightness and indistinguishability},
  author={Liu, Jin and Su, Rongbin and Wei, Yuming and Yao, Beimeng and Silva, Saimon Filipe Covre da and Yu, Ying and Iles-Smith, Jake and Srinivasan, Kartik and Rastelli, Armando and Li, Juntao and others},
  journal={Nature nanotechnology},
  volume={14},
  number={6},
  pages={586--593},
  year={2019},
  publisher={Nature Publishing Group UK London}
}

@article{schimpf2021quantum,
  title={Quantum dots as potential sources of strongly entangled photons: Perspectives and challenges for applications in quantum networks},
  author={Schimpf, Christian and Reindl, Marcus and Basso Basset, Francesco and J{\"o}ns, Klaus D and Trotta, Rinaldo and Rastelli, Armando},
  journal={Applied Physics Letters},
  volume={118},
  number={10},
  year={2021},
  publisher={AIP Publishing}
}

@article{strobel2024quantum,
	title = {Telecom-wavelength quantum teleportation using frequency-converted photons from remote quantum dots},
	volume = {16},
	issn = {2041-1723},
	url = {https://www.nature.com/articles/s41467-025-65912-8},
	doi = {10.1038/s41467-025-65912-8},
	number = {1},
	urldate = {2025-11-17},
	journal = {Nature Communications},
	author = {Strobel, Tim and Vyvlecka, Michal and Neureuther, Ilenia and Bauer, Tobias and Schäfer, Marlon and Kazmaier, Stefan and Sharma, Nand Lal and Joos, Raphael and Weber, Jonas H. and Nawrath, Cornelius and Nie, Weijie and Bhayani, Ghata and Hopfmann, Caspar and Becher, Christoph and Michler, Peter and Portalupi, Simone Luca},
	month = nov,
	year = {2025},
	pages = {10027},
}

@article{lu2021quantum,
  title={Quantum-dot single-photon sources for the quantum internet},
  author={Lu, Chao-Yang and Pan, Jian-Wei},
  journal={Nature Nanotechnology},
  volume={16},
  number={12},
  pages={1294--1296},
  year={2021},
  publisher={Nature Publishing Group UK London}
}

@article{laneve2024quantum,
	title = {Quantum teleportation with dissimilar quantum dots over a hybrid quantum network},
	volume = {16},
	issn = {2041-1723},
	url = {https://www.nature.com/articles/s41467-025-65911-9},
	doi = {10.1038/s41467-025-65911-9},
	number = {1},
	urldate = {2025-11-17},
	journal = {Nature Communications},
	author = {Laneve, Alessandro and Ronco, Giuseppe and Beccaceci, Mattia and Barigelli, Paolo and Salusti, Francesco and Claro-Rodriguez, Nicolas and De Pascalis, Giorgio and Suprano, Alessia and Chiaudano, Leone and Sch{\"o}ll, Eva and Hanschke, Lukas and Krieger, Tobias M. and Buchinger, Quirin and Covre Da Silva, Saimon F. and Neuwirth, Julia and Stroj, Sandra and Höfling, Sven and Huber-Loyola, Tobias and Usuga Castaneda, Mario A. and Carvacho, Gonzalo and Spagnolo, Nicolò and Rota, Michele B. and Basso Basset, Francesco and Rastelli, Armando and Sciarrino, Fabio and Jöns, Klaus D. and Trotta, Rinaldo},
	month = nov,
	year = {2025},
	pages = {10028},
}

@article{chen2013coherent,
  title={Coherent optical memory with high storage efficiency and large fractional delay},
  author={Chen, Yi-Hsin and Lee, Meng-Jung and Wang, I-Chung and Du, Shengwang and Chen, Yong-Fan and Chen, Ying-Cheng and Yu, Ite A},
  journal={Physical review letters},
  volume={110},
  number={8},
  pages={083601},
  year={2013},
  publisher={APS}
}

@article{namazi2017ultralow,
  title={Ultralow-noise room-temperature quantum memory for polarization qubits},
  author={Namazi, Mehdi and Kupchak, Connor and Jordaan, Bertus and Shahrokhshahi, Reihaneh and Figueroa, Eden},
  journal={Physical Review Applied},
  volume={8},
  number={3},
  pages={034023},
  year={2017},
  publisher={APS}
}

@article{tang2015storage,
  title={Storage of multiple single-photon pulses emitted from a quantum dot in a solid-state quantum memory},
  author={Tang, Jian-Shun and Zhou, Zong-Quan and Wang, Yi-Tao and Li, Yu-Long and Liu, Xiao and Hua, Yi-Lin and Zou, Yang and Wang, Shuang and He, De-Yong and Chen, Geng and others},
  journal={Nature communications},
  volume={6},
  number={1},
  pages={8652},
  year={2015},
  publisher={Nature Publishing Group UK London}
}

@article{wolters2017simple,
  title = {Simple Atomic Quantum Memory Suitable for Semiconductor Quantum Dot Single Photons},
  author = {Wolters, Janik and Buser, Gianni and Horsley, Andrew and B\'eguin, Lucas and J\"ockel, Andreas and Jahn, Jan-Philipp and Warburton, Richard J. and Treutlein, Philipp},
  journal = {Phys. Rev. Lett.},
  volume = {119},
  issue = {6},
  pages = {060502},
  numpages = {5},
  year = {2017},
  month = {Aug},
  publisher = {American Physical Society},
  doi = {10.1103/PhysRevLett.119.060502},
  url = {https://link.aps.org/doi/10.1103/PhysRevLett.119.060502}
}

@article{neuwirth2022multipair,
  title = {Multipair-free source of entangled photons in the solid state},
  volume = {106},
  ISSN = {2469-9969},
  url = {http://dx.doi.org/10.1103/PhysRevB.106.L241402},
  DOI = {10.1103/physrevb.106.l241402},
  number = {24},
  journal = {Physical Review B},
  publisher = {American Physical Society (APS)},
  author = {Neuwirth,  Julia and Basso Basset,  Francesco and Rota,  Michele B. and Hartel,  Jan-Gabriel and Sartison,  Marc and Covre da Silva,  Saimon F. and J\"{o}ns,  Klaus D. and Rastelli,  Armando and Trotta,  Rinaldo},
  year = {2022},
  month = dec 
}

@article{hanschke2018quantum,
  title = {Quantum dot single-photon sources with ultra-low multi-photon probability},
  volume = {4},
  ISSN = {2056-6387},
  url = {http://dx.doi.org/10.1038/s41534-018-0092-0},
  DOI = {10.1038/s41534-018-0092-0},
  number = {1},
  journal = {npj Quantum Information},
  publisher = {Springer Science and Business Media LLC},
  author = {Hanschke,  Lukas and Fischer,  Kevin A. and Appel,  Stefan and Lukin,  Daniil and Wierzbowski,  Jakob and Sun,  Shuo and Trivedi,  Rahul and Vučković,  Jelena and Finley,  Jonathan J. and M\"{u}ller,  Kai},
  year = {2018},
  month = sep 
}

@article{huber2018strain,
  title = {Strain-Tunable GaAs Quantum Dot: A Nearly Dephasing-Free Source of Entangled Photon Pairs on Demand},
  volume = {121},
  ISSN = {1079-7114},
  url = {http://dx.doi.org/10.1103/PhysRevLett.121.033902},
  DOI = {10.1103/physrevlett.121.033902},
  number = {3},
  journal = {Physical Review Letters},
  publisher = {American Physical Society (APS)},
  author = {Huber,  Daniel and Reindl,  Marcus and Covre da Silva,  Saimon Filipe and Schimpf,  Christian and Martín-Sánchez,  Javier and Huang,  Huiying and Piredda,  Giovanni and Edlinger,  Johannes and Rastelli,  Armando and Trotta,  Rinaldo},
  year = {2018},
  month = jul 
}

@article{schimpf2023hyperfine,
  title = {Hyperfine interaction limits polarization entanglement of photons from semiconductor quantum dots},
  volume = {108},
  ISSN = {2469-9969},
  url = {http://dx.doi.org/10.1103/PhysRevB.108.L081405},
  DOI = {10.1103/physrevb.108.l081405},
  number = {8},
  journal = {Physical Review B},
  publisher = {American Physical Society (APS)},
  author = {Schimpf,  Christian and Basso Basset,  Francesco and Aigner,  Maximilian and Attenender,  Wolfgang and Ginés,  Laia and Undeutsch,  Gabriel and Reindl,  Marcus and Huber,  Daniel and Gangloff,  Dorian and Chekhovich,  Evgeny A. and Schneider,  Christian and H\"{o}fling,  Sven and Predojević,  Ana and Trotta,  Rinaldo and Rastelli,  Armando},
  year = {2023},
  month = aug 
}

@article{rota2024source,
  title={A source of entangled photons based on a cavity-enhanced and strain-tuned \ce{GaAs} quantum dot},
  author={Rota, Michele B and Krieger, Tobias M and Buchinger, Quirin and Beccaceci, Mattia and Neuwirth, Julia and Huet, H{\^e}lio and Horov{\'a}, Nikola and Lovicu, Gabriele and Ronco, Giuseppe and Covre da Silva, Saimon F and others},
  journal={eLight},
  volume={4},
  number={1},
  pages={13},
  year={2024},
  publisher={Springer}
}

@article{altepeter2005photonic,
  title={Photonic state tomography},
  author={Altepeter, Joseph B and Jeffrey, Evan R and Kwiat, Paul G},
  journal={Advances in atomic, molecular, and optical physics},
  volume={52},
  pages={105--159},
  year={2005},
  publisher={Elsevier}
}

@article{simon2005creating,
  title={Creating single time-bin-entangled photon pairs},
  author={Simon, Christoph and Poizat, Jean-Philippe},
  journal={Physical review letters},
  volume={94},
  number={3},
  pages={030502},
  year={2005},
  publisher={APS}
}

@article{rota2020entanglement,
  title={Entanglement teleportation with photons from quantum dots: toward a solid-state based quantum network},
  author={Rota, Michele B and Basso Basset, Francesco and Tedeschi, Davide and Trotta, Rinaldo},
  journal={IEEE Journal of Selected Topics in Quantum Electronics},
  volume={26},
  number={3},
  pages={1--16},
  year={2020},
  publisher={IEEE}
}

@article{mateo2015demystifying,
author = {Martinez-Mateo, Jesus and Pacher, Christoph and Peev, Momtchil and Ciurana, Alex and Martin, Vicente},
title = {Demystifying the information reconciliation protocol cascade},
year = {2015},
issue_date = {April 2015},
publisher = {Rinton Press, Incorporated},
address = {Paramus, NJ},
volume = {15},
number = {5–6},
issn = {1533-7146},
journal = {Quantum Info. Comput.},
month = apr,
pages = {453–477},
numpages = {25}
}

@article{schimpf2021entanglement,
  title={Entanglement-based quantum key distribution with a blinking-free quantum dot operated at a temperature up to 20 K},
  author={Schimpf, Christian and Manna, Santanu and Covre da Silva, Saimon F and Aigner, Maximilian and Rastelli, Armando},
  journal={Advanced Photonics},
  volume={3},
  number={6},
  pages={065001--065001},
  year={2021},
  publisher={Society of Photo-Optical Instrumentation Engineers}
}

@article{lutkenhaus1999bell,
  title={{B}ell measurements for teleportation},
  author={L{\"u}tkenhaus, Norbert and Calsamiglia, John and Suominen, K-A},
  journal={Physical Review A},
  volume={59},
  number={5},
  pages={3295},
  year={1999},
  publisher={APS}
}

@article{huber2013measurement,
  title={Measurement and modification of biexciton-exciton time correlations},
  author={Huber, Tobias and Predojevi{\'c}, Ana and Zoubi, Hashem and Jayakumar, Harishankar and Solomon, Glenn S and Weihs, Gregor},
  journal={Optics Express},
  volume={21},
  number={8},
  pages={9890--9898},
  year={2013},
  publisher={Optical Society of America}
}

@article{briegel1998quantum,
  title = {Quantum Repeaters: The Role of Imperfect Local Operations in Quantum Communication},
  author = {Briegel, H.-J. and D\"ur, W. and Cirac, J. I. and Zoller, P.},
  journal = {Phys. Rev. Lett.},
  volume = {81},
  issue = {26},
  pages = {5932--5935},
  numpages = {0},
  year = {1998},
  month = {Dec},
  publisher = {American Physical Society},
  doi = {10.1103/PhysRevLett.81.5932},
  url = {https://link.aps.org/doi/10.1103/PhysRevLett.81.5932}
}

@article{hosak2021effect,
  title = {Effect of source statistics on utilizing photon entanglement in quantum key distribution},
  volume = {103},
  ISSN = {2469-9934},
  url = {http://dx.doi.org/10.1103/PhysRevA.103.042411},
  DOI = {10.1103/physreva.103.042411},
  number = {4},
  journal = {Physical Review A},
  publisher = {American Physical Society (APS)},
  author = {Hošák,  Radim and Straka,  Ivo and Predojević,  Ana and Filip,  Radim and Ježek,  Miroslav},
  year = {2021},
  month = apr 
}

@article{jayakumar2013deterministic,
  title = {Deterministic Photon Pairs and Coherent Optical Control of a Single Quantum Dot},
  author = {Jayakumar, Harishankar and Predojevi\ifmmode \acute{c}\else \'{c}\fi{}, Ana and Huber, Tobias and Kauten, Thomas and Solomon, Glenn S. and Weihs, Gregor},
  journal = {Phys. Rev. Lett.},
  volume = {110},
  issue = {13},
  pages = {135505},
  numpages = {5},
  year = {2013},
  month = {Mar},
  publisher = {American Physical Society},
  doi = {10.1103/PhysRevLett.110.135505},
  url = {https://link.aps.org/doi/10.1103/PhysRevLett.110.135505}
}

@article{muller2014ondemand,
  title = {On-demand generation of indistinguishable polarization-entangled photon pairs},
  volume = {8},
  ISSN = {1749-4893},
  url = {http://dx.doi.org/10.1038/NPHOTON.2013.377},
  DOI = {10.1038/nphoton.2013.377},
  number = {3},
  journal = {Nature Photonics},
  publisher = {Springer Science and Business Media LLC},
  author = {M\"{u}ller,  M. and Bounouar,  S. and J\"{o}ns,  K. D. and Gl\"{a}ssl,  M. and Michler,  P.},
  year = {2014},
  month = feb,
  pages = {224–228}
}

@article{ling2008experimental,
  title = {Experimental quantum key distribution based on a {B}ell test},
  volume = {78},
  ISSN = {1094-1622},
  url = {http://dx.doi.org/10.1103/PhysRevA.78.020301},
  DOI = {10.1103/physreva.78.020301},
  number = {2},
  journal = {Physical Review A},
  publisher = {American Physical Society (APS)},
  author = {Ling,  Alexander and Peloso,  Matthew P. and Marcikic,  Ivan and Scarani,  Valerio and Lamas-Linares,  Antía and Kurtsiefer,  Christian},
  year = {2008},
  month = aug 
}

@article{clauser1969proposed,
  title = {Proposed Experiment to Test Local Hidden-Variable Theories},
  author = {Clauser, John F. and Horne, Michael A. and Shimony, Abner and Holt, Richard A.},
  journal = {Phys. Rev. Lett.},
  volume = {23},
  issue = {15},
  pages = {880--884},
  numpages = {0},
  year = {1969},
  month = {Oct},
  publisher = {American Physical Society},
  doi = {10.1103/PhysRevLett.23.880},
  url = {https://link.aps.org/doi/10.1103/PhysRevLett.23.880}
}

@article{pan1998experimental,
  title = {Experimental Entanglement Swapping: Entangling Photons That Never Interacted},
  author = {Pan, Jian-Wei and Bouwmeester, Dik and Weinfurter, Harald and Zeilinger, Anton},
  journal = {Phys. Rev. Lett.},
  volume = {80},
  issue = {18},
  pages = {3891--3894},
  numpages = {0},
  year = {1998},
  month = {May},
  publisher = {American Physical Society},
  doi = {10.1103/PhysRevLett.80.3891},
  url = {https://link.aps.org/doi/10.1103/PhysRevLett.80.3891}
}

@article{basso2019entanglement,
  title={Entanglement swapping with photons generated on demand by a quantum dot},
  author={Basso Basset, Francesco and Rota, Michele B and Schimpf, Christian and Tedeschi, Davide and Zeuner, Katharina D and Covre da Silva, Saimon F and Reindl, Marcus and Zwiller, Val and J{\"o}ns, Klaus D and Rastelli, Armando and others},
  journal={Physical Review Letters},
  volume={123},
  number={16},
  pages={160501},
  year={2019},
  publisher={APS}
}

@article{zopf2019entanglement,
  title={Entanglement swapping with semiconductor-generated photons violates {B}ell’s inequality},
  author={Zopf, Michael and Keil, Robert and Chen, Yan and Yang, Jingzhong and Chen, Disheng and Ding, Fei and Schmidt, Oliver G},
  journal={Physical review letters},
  volume={123},
  number={16},
  pages={160502},
  year={2019},
  publisher={APS}
}

@article{basso2021quantum,
  title={Quantum teleportation with imperfect quantum dots},
  author={Basso Basset, Francesco and Salusti, Francesco and Schweickert, Lucas and Rota, Michele B and Tedeschi, Davide and Covre da Silva, Saimon F and Roccia, Emanuele and Zwiller, Val and J{\"o}ns, Klaus D and Rastelli, Armando and others},
  journal={npj Quantum Information},
  volume={7},
  number={1},
  pages={7},
  year={2021},
  publisher={Nature Publishing Group UK London}
}

@article{zhai2022quantum,
  title={Quantum interference of identical photons from remote GaAs quantum dots},
  author={Zhai, Liang and Nguyen, Giang N and Spinnler, Clemens and Ritzmann, Julian and L{\"o}bl, Matthias C and Wieck, Andreas D and Ludwig, Arne and Javadi, Alisa and Warburton, Richard J},
  journal={Nature nanotechnology},
  volume={17},
  number={8},
  pages={829--833},
  year={2022},
  publisher={Nature Publishing Group UK London}
}

@article{scholl2020crux,
  title={Crux of using the cascaded emission of a three-level quantum ladder system to generate indistinguishable photons},
  author={Sch{\"o}ll, Eva and Schweickert, Lucas and Hanschke, Lukas and Zeuner, Katharina D and Sbresny, Friedrich and Lettner, Thomas and Trivedi, Rahul and Reindl, Marcus and Covre da Silva, Saimon Filipe and Trotta, Rinaldo and others},
  journal={Physical Review Letters},
  volume={125},
  number={23},
  pages={233605},
  year={2020},
  publisher={APS}
}

@article{legero2004quantum,
  title={Quantum beat of two single photons},
  author={Legero, Thomas and Wilk, Tatjana and Hennrich, Markus and Rempe, Gerhard and Kuhn, Axel},
  journal={Physical review letters},
  volume={93},
  number={7},
  pages={070503},
  year={2004},
  publisher={APS}
}

@article{acin2006efficient,
  title={Efficient quantum key distribution secure against no-signalling eavesdroppers},
  author={Acin, Antonio and Massar, Serge and Pironio, Stefano},
  journal={New Journal of Physics},
  volume={8},
  number={8},
  pages={126},
  year={2006},
  publisher={IOP Publishing}
}

@article{acin2007device,
  title={Device-independent security of quantum cryptography against collective attacks},
  author={Ac{\'\i}n, Antonio and Brunner, Nicolas and Gisin, Nicolas and Massar, Serge and Pironio, Stefano and Scarani, Valerio},
  journal={Physical Review Letters},
  volume={98},
  number={23},
  pages={230501},
  year={2007},
  publisher={APS}
}

@article{pironio2009device,
  title={Device-independent quantum key distribution secure against collective attacks},
  author={Pironio, Stefano and Ac{\'\i}n, Antonio and Brunner, Nicolas and Gisin, Nicolas and Massar, Serge and Scarani, Valerio},
  journal={New Journal of Physics},
  volume={11},
  number={4},
  pages={045021},
  year={2009},
  publisher={IOP Publishing}
}

@article{davis2025entanglement,
  title={Entanglement swapping systems toward a quantum internet},
  author={Davis, Samantha I and Valivarthi, Raju and Cameron, Andrew and Pena, Cristian and Xie, Si and Narvaez, Lautaro and Lauk, Nikolai and Li, Chang and Taylor, Kelsie and Youssef, Rahaf and others},
  journal={arXiv preprint arXiv:2503.18906},
  year={2025}
}

@article{ma2007quantum,
  title={Quantum key distribution with entangled photon sources},
  author={Ma, Xiongfeng and Fung, Chi-Hang Fred and Lo, Hoi-Kwong},
  journal={Physical Review A—Atomic, Molecular, and Optical Physics},
  volume={76},
  number={1},
  pages={012307},
  year={2007},
  publisher={APS}
}

@article{bratzik2013quantum,
  title={Quantum repeaters and quantum key distribution: The impact of entanglement distillation on the secret key rate},
  author={Bratzik, Sylvia and Abruzzo, Silvestre and Kampermann, Hermann and Bru{\ss}, Dagmar},
  journal={Physical Review A—Atomic, Molecular, and Optical Physics},
  volume={87},
  number={6},
  pages={062335},
  year={2013},
  publisher={APS}
}

@article{dur1999quantum,
  title = {Quantum repeaters based on entanglement purification},
  author = {D\"ur, W. and Briegel, H.-J. and Cirac, J. I. and Zoller, P.},
  journal = {Phys. Rev. A},
  volume = {59},
  issue = {1},
  pages = {169--181},
  numpages = {0},
  year = {1999},
  publisher = {APS}
}

@article{deutsch1996quantum,
  title = {Quantum Privacy Amplification and the Security of Quantum Cryptography over Noisy Channels},
  author = {Deutsch, David and Ekert, Artur and Jozsa, Richard and Macchiavello, Chiara and Popescu, Sandu and Sanpera, Anna},
  journal = {Phys. Rev. Lett.},
  volume = {77},
  issue = {13},
  pages = {2818--2821},
  numpages = {0},
  year = {1996},
  publisher = {APS}
}

@article{zhang2022device,
  title={A device-independent quantum key distribution system for distant users},
  author={Zhang, Wei and van Leent, Tim and Redeker, Kai and Garthoff, Robert and Schwonnek, Ren{\'e} and Fertig, Florian and Eppelt, Sebastian and Rosenfeld, Wenjamin and Scarani, Valerio and Lim, Charles C-W and others},
  journal={Nature},
  volume={607},
  number={7920},
  pages={687--691},
  year={2022},
  publisher={Nature Publishing Group UK London}
}

@article{nadlinger2022experimental,title={Experimental quantum key distribution certified by {B}ell's theorem},author={Nadlinger, David P and Drmota, Peter and Nichol, Bethan C and Araneda, Gabriel and Main, Dougal and Srinivas, Raghavendra and Lucas, David M and Ballance, Christopher J and Ivanov, Kirill and Tan, EY-Z and others},
  journal={Nature},
  volume={607},
  number={7920},
  pages={682--686},
  year={2022},
  publisher={Nature Publishing Group UK London}
}

@article{undeutsch2025electric,
  title={Electric-field control of photon indistinguishability in cascaded decays in quantum dots},
  author={Undeutsch, Gabriel and Aigner, Maximilian and Garcia Jr, Ailton J and Reindl, Johannes and Peter, Melina and Mader, Simon and Weidinger, Christian and Covre da Silva, Saimon F and Manna, Santanu and Sch{\"o}ll, Eva and others},
  journal={Nano Letters},
  volume={25},
  number={17},
  pages={7121--7127},
  year={2025},
  publisher={ACS Publications}
}

@misc{gumberidze2025feasibilityentanglementbasedqkdprotocols,
      title={Feasibility of entanglement-based QKD protocols with SPDC and QD sources}, 
      author={Mariia Gumberidze and Vladyslav C. Usenko},
      year={2025},
      eprint={2510.20528},
      archivePrefix={arXiv},
      primaryClass={quant-ph},
      url={https://arxiv.org/abs/2510.20528}, 
}

@article{beccaceci_all-photonic_2025,
	title = {All-photonic entanglement swapping with remote quantum dots},
	author = {Beccaceci, Mattia and Ronco, Giuseppe and Cienzo, Fabrizio and Bassetti, Pierpaolo and Laneve, Alessandro and Basso Basset, Francesco  and Krieger, Tobias M and Buchinger, Qurin and Salusti, Francesco and Souza, Barbara and Kuhn, Silke and Huber-Loyola, Tobias and Rastelli, Armando and Rota, Michele B and Trotta, Rinaldo},
    journal = {arXiv preprint},
    year={2025},
    eprint={2512.10651},
    archivePrefix={arXiv},
    primaryClass={quant-ph},
    url={https://arxiv.org/abs/2512.10651}, 
}

@article{zapatero2023advances,
  title={Advances in device-independent quantum key distribution},
  author={Zapatero, V{\'\i}ctor and van Leent, Tim and Arnon-Friedman, Rotem and Liu, Wen-Zhao and Zhang, Qiang and Weinfurter, Harald and Curty, Marcos},
  journal={npj quantum information},
  volume={9},
  number={1},
  pages={10},
  year={2023},
  publisher={Nature Publishing Group UK London}
}

@article{lu2026device,
  title={Device-independent quantum key distribution over 100 km with single atoms},
  author={Lu, Bo-Wei and Yang, Chao-Wei and Wang, Run-Qi and Gao, Bo-Feng and Zhen, Yi-Zheng and Wang, Zhen-Gang and Shi, Jia-Kai and Ren, Zhong-Qi and Hahn, Thomas A and Tan, Ernest Y-Z and others},
  journal={Science},
  volume={391},
  number={6785},
  pages={592--597},
  year={2026},
  publisher={American Association for the Advancement of Science}
}

@article{liu2026long,
  title={Long-lived remote ion-ion entanglement for scalable quantum repeaters},
  author={Liu, Wen-Zhao and Zhou, Ya-Bin and Chen, Jiu-Peng and Wang, Bin and Teng, Ao and Han, Xiao-Wen and Liu, Guang-Cheng and Zhang, Zhi-Jiong and Yang, Yi and Liu, Feng-Guang and others},
  journal={Nature},
  pages={1--3},
  year={2026},
  publisher={Nature Publishing Group UK London}
}

@article{mantri2025comparing,
  title={Comparing one-and two-way quantum repeater architectures},
  author={Mantri, Prateek and Goodenough, Kenneth and Towsley, Don},
  journal={Communications Physics},
  volume={8},
  number={1},
  pages={300},
  year={2025},
  publisher={Nature Publishing Group UK London}
}

@article{goodenough2021optimizing,
  title={Optimizing repeater schemes for the quantum internet},
  author={Goodenough, Kenneth and Elkouss, David and Wehner, Stephanie},
  journal={Physical Review A},
  volume={103},
  number={3},
  pages={032610},
  year={2021},
  publisher={APS}
}

@article{jiang2026entanglement,
  title={Entanglement Swapping Enables the Practical Security of Quantum Cryptography},
  author={Jiang, Yang-Fan and Huang, Liang and Zhang, Yu-Zhe and Zhang, Likang and Sun, Qi-Chao and Li, Zheng-Ping and Li, Hao and Zhang, Weijun and You, Lixing and Xu, Feihu and others},
  journal={Entropy},
  volume={28},
  number={5},
  pages={518},
  year={2026},
  publisher={MDPI}
}

@article{patel2010two,
  title={Two-photon interference of the emission from electrically tunable remote quantum dots},
  author={Patel, Raj B and Bennett, Anthony J and Farrer, Ian and Nicoll, Christine A and Ritchie, David A and Shields, Andrew J},
  journal={Nature photonics},
  volume={4},
  number={9},
  pages={632--635},
  year={2010},
  publisher={Nature Publishing Group UK London}
}

@article{gisin2010proposal,
  title={Proposal for Implementing Device-Independent Quantum Key Distribution Based on a Heralded Qubit Amplifier},
  author={Gisin, Nicolas and Pironio, Stefano and Sangouard, Nicolas},
  journal={Physical review letters},
  volume={105},
  number={7},
  pages={070501},
  year={2010},
  publisher={APS}
}

@article{kolodynski2020device,
  title={Device-independent quantum key distribution with single-photon sources},
  author={Ko{\l}ody{\'n}ski, Jan and M{\'a}ttar, Alejandro and Skrzypczyk, Paul and Woodhead, Erik and Cavalcanti, Daniel and Banaszek, Konrad and Ac{\'\i}n, Antonio},
  journal={Quantum},
  volume={4},
  pages={260},
  year={2020},
  publisher={Verein zur F{\"o}rderung des Open Access Publizierens in den Quantenwissenschaften}
}

@article{patel2008postselective,
  title={Postselective Two-Photon Interference from a Continuous Nonclassical Stream of Photons Emitted by a Quantum Dot},
  author={Patel, RB and Bennett, AJ and Cooper, K and Atkinson, P and Nicoll, CA and Ritchie, DA and Shields, AJ},
  journal={Physical Review Letters},
  volume={100},
  number={20},
  pages={207405},
  year={2008},
  publisher={APS}
}

@article{anderson2020quantum,
  title={Quantum teleportation using highly coherent emission from telecom C-band quantum dots},
  author={Anderson, M and M{\"u}ller, T and Huwer, J and Skiba-Szymanska, J and Krysa, AB and Stevenson, RM and Heffernan, J and Ritchie, DA and Shields, AJ},
  journal={npj Quantum Information},
  volume={6},
  number={1},
  pages={14},
  year={2020},
  publisher={Nature Publishing Group UK London}
}

@article{huwer2017quantum,
  title={Quantum-dot-based telecommunication-wavelength quantum relay},
  author={Huwer, J and Stevenson, RM and Skiba-Szymanska, J and Ward, MB and Shields, AJ and Felle, M and Farrer, I and Ritchie, David Alastair and Penty, Richard Vincent},
  journal={Physical Review Applied},
  volume={8},
  number={2},
  pages={024007},
  year={2017},
  publisher={APS}
}

@article{anderson2020gigahertz,
  title={Gigahertz-clocked teleportation of time-bin qubits with a quantum dot in the telecommunication C band},
  author={Anderson, M and M{\"u}ller, T and Skiba-Szymanska, J and Krysa, AB and Huwer, J and Stevenson, RM and Heffernan, J and Ritchie, DA and Shields, AJ},
  journal={Physical Review Applied},
  volume={13},
  number={5},
  pages={054052},
  year={2020},
  publisher={APS}
}

@article{ikuta2018polarization,
  title={Polarization insensitive frequency conversion for an atom-photon entanglement distribution via a telecom network},
  author={Ikuta, Rikizo and Kobayashi, Toshiki and Kawakami, Tetsuo and Miki, Shigehito and Yabuno, Masahiro and Yamashita, Taro and Terai, Hirotaka and Koashi, Masato and Mukai, Tetsuya and Yamamoto, Takashi and others},
  journal={Nature communications},
  volume={9},
  number={1},
  pages={1997},
  year={2018},
  publisher={Nature Publishing Group UK London}
}

@article{weber2019two,
  title={Two-photon interference in the telecom C-band after frequency conversion of photons from remote quantum emitters},
  author={Weber, Jonas H and Kambs, Benjamin and Kettler, Jan and Kern, Simon and Maisch, Julian and Vural, H{\"u}seyin and Jetter, Michael and Portalupi, Simone L and Becher, Christoph and Michler, Peter},
  journal={Nature nanotechnology},
  volume={14},
  number={1},
  pages={23--26},
  year={2019},
  publisher={Nature Publishing Group UK London}
}

@article{bourgoin_comprehensive_2013,
	title = {A comprehensive design and performance analysis of low {Earth} orbit satellite quantum communication},
	volume = {15},
	copyright = {http://iopscience.iop.org/info/page/text-and-data-mining},
	issn = {1367-2630},
	url = {https://iopscience.iop.org/article/10.1088/1367-2630/15/2/023006},
	doi = {10.1088/1367-2630/15/2/023006},
	number = {2},
	urldate = {2026-04-18},
	journal = {New Journal of Physics},
	author = {Bourgoin, J-P and Meyer-Scott, E and Higgins, B L and Helou, B and Erven, C and Hübel, H and Kumar, B and Hudson, D and D'Souza, I and Girard, R and Laflamme, R and Jennewein, T},
	month = feb,
	year = {2013},
	pages = {023006},
}

\end{document}